\documentclass[onecolumn]{aa}

\usepackage{graphicx}
\usepackage{txfonts}
\usepackage{url}            % simple URL typesetting
\usepackage{booktabs}       % professional-quality tables
\usepackage{amsfonts}       % blackboard math symbols
\usepackage{nicefrac}       % compact symbols for 1/2, etc.
\usepackage{microtype}      % microtypography
\usepackage{subcaption}
\usepackage{caption}
\usepackage{graphicx}
\usepackage{tabularx}

\begin{document} 
\nolinenumbers
   \title{Strategy for sensing petal mode in presence of AO residual turbulence with pyramid wavefront sensor}

   \author{Nicolas Levraud \inst{1,2,3}
            \and
          	Vincent Chambouleyron \inst{4}
           \and
           Jean François Sauvage \inst{1,3}
           \and
           Benoit Neichel \inst{3}
           \and
           Mahawa Cisse \inst{1,3}
           \and
           Olivier Fauvarque \inst{5}
           \and
           Guido Agapito \inst{2}
           \and
           Cédric Plantet \inst{2}
           \and
           Anne Laure Cheffot \inst{2}
           \and
           Enrico Pinna \inst{2}
           \and
           Simone Esposito \inst{2}
                 \and
           Thierry Fusco \inst{1,3}
    }

   \institute{DOTA, ONERA, Université Paris Saclay, F-91123 Palaiseau, France
                \email{nicolas.levraud@obspm.fr}
                \and
                INAF - Osservatorio Astronomico di Arcetri, 50125 Firenze FI, Italie
                \and
                Aix Marseille Univ, CNRS, CNES, LAM, 13013 Marseille, France
                \and
                University of California Santa Cruz, 1156 High St, Santa Cruz, USA
                \and   
                IFREMER, Laboratoire Detection, Capteurs et Mesures (LDCM), Centre Bretagne, ZI de la Pointe du Diable, CS 10070, 29280, Plouzane, France
             }

   %\date{Received September 15, 1996; accepted March 16, 1997}

% \abstract{}{}{}{}{} 
% 5 {} token are mandatory
 
  \abstract
   {With the Extremely Large Telescope-generation telescopes come new challenges. The complexity of these telescopes' pupil creates new problems for Adaptive Optics, which prevent the telescope from reaching the theoretical resolution that their size shall allow. In particular, the large spiders necessary to support the massive optics of these telescopes create discontinuities in the wavefront measurement. These discontinuities appear as a new phase error dubbed the `petal mode'. This error is described as a differential piston between the fragment of the pupil separated by the spiders and is responsible for a strong degradation of the imaging quality,  reducing the European Extremely Large Telescope's (ELT) resolution to a 15m telescope resolution. }
   {The aim of this paper is to study the measurement of the petal mode by adaptive optics sensors. In particular, we want to understand why the Pyramid Wavefront Sensor (PyWFS), the first light wavefront sensor of all ELT-generation telescopes, cannot measure this petal mode under normal conditions and how to allow this measurement by adapting the Adaptive optics control scheme and the PyWFS.}
   {To facilitate our study, we consider a simplified version of the petal mode, featuring a simpler pupil than the ELT. This allows us to quickly simulate the properties of the petal mode and its measurement by the PyWFS. We studied specifically how a system that separates the atmospheric turbulence from the petal measurement would behave. Studying the petal mode's power spectral density, we propose to use a spatial filter to reduce the contribution of AO residuals to the benefit of petal mode contribution, eventually helping its measurement. Finally, we demonstrate our proposed system with end-to-end simulations. }
   {A solution proposed to measure the petal mode is to use an unmodulated PyWFS (uPyWFS) but the uPyWFS does not make accurate measurements in the presence of atmospheric residuals. A spatial filtering step, consisting of a pinhole around the pyramid tip, reduces the first path residuals seen by the uPyWFS and restores its accuracy. This system was able to measure and control petal mode during the end-to-end simulation.}
   {To address the petal problem, a two-path adaptive optics with a sensor dedicated to the measurement of the petal mode seems necessary. The question remains as to what could be used as the second path petalometer. Through this paper, we demonstrate that an uPyWFS can confuse the petal mode with the residuals from the first path. However, adding a spatial filter on top of said uPyWFS makes it a good petalometer candidate. This spatial filtering step makes the uPyWFS less sensitive to the first path residuals while retaining its ability to measure the petal mode.}


   \maketitle
%
%-------------------------------------------------------------------
\nolinenumbers
\section{Introduction}

%\subsection{ELT class telescopes}
Due to the atmospheric turbulence, building a telescope bigger than $15$ centimeters in visible light improves only in its capacity to collect more light, but not in the angular resolution. Telescopes bigger than this threshold have to use Adaptive Optics (AO) to compensate for the effect of the atmosphere before the light reaches the science instruments. This AO compensation step allows one to restore the diffraction limit and the gain in resolution offered by a larger primary pupil. The next generation of 30m-class telescopes is under construction and will have AO capability from the first light. In particular, in this paper, we focus on the Extremely Large Telescope (ELT)  (\cite{cayrelEELTOptomechanicsOverview2012}).

%\subsection{PyWFS Wavefront sensor}
To meet the challenge of the size of these new telescopes, the classical AO wavefront sensor, the Shack-Hartmann (SH), is less adapted. Its separation of the pupil in multiple subpupils makes it less sensitive for each subpupil, as shown by \cite{verinaudNatureMeasurementsProvided2004a}. To produce a diffraction-limited Point Spread function (PSF) on a larger telescope, the number of actuators needed, and consequently, the number of subpupils required for the measurement increase. The SH becomes more susceptible to noise and hence not adapted to low flux regimes. Furthermore, the SH needs more pixels to the point where it is not compatible with current AO cameras. As a result, the SH is being replaced by a new kind of WaveFront Sensor (WFS) for Single Conjugated Adaptive Optics (SCAO), where our only source of light is a Natural Guide Star (NGS). This kind of WFS is called the Pyramid Wavefront Sensor (PyWFS), proposed by \cite{ragazzoniPupilPlaneWavefront1996a}, and is employed in the SCAO mode of each 30m-class telescope.

%\subsection{Science case}
The utility of SCAO is to provide high-quality wavefront correction near a bright natural guide star. Specifically, the most demanding science case for such a system is the search for exoplanets, which requires high AO performance to detect faint objects close to their stars. The PyWFS allows the ELT-class telescopes to reach their diffraction limit and angular resolution with fainter targets than a SH.

%\subsection{Petal mode} 

Simulations of the HARMONI SCAO module taking the ELT pupil and the segmented M4 (\cite{schwartzSensingControlSegmented2017} ) showed a new limit to the SCAO performances, with differential piston appearing between the pupil fragment separated by the large spiders of the ELT.  This is due to the inability of the SCAO system to measure this differential piston and the absence of constraints on this differential piston with a segmented DM. Owing to its occurrence in multiple systems, this effect has been specifically termed the `petal mode basis', which is a linear combination of the differential pistons between the pupil fragments. This effect can be greatly reduced by adding constraints on the ELT deformable mirror, using techniques such as minioning (originally called slaving) or a continual DM basis \cite{bertrou-cantouPetalometryELTDealing2020}.

\begin{figure}
\centering
    \begin{subfigure}[t]{0.45\columnwidth}
        \centering
        \includegraphics[width=0.95\columnwidth]{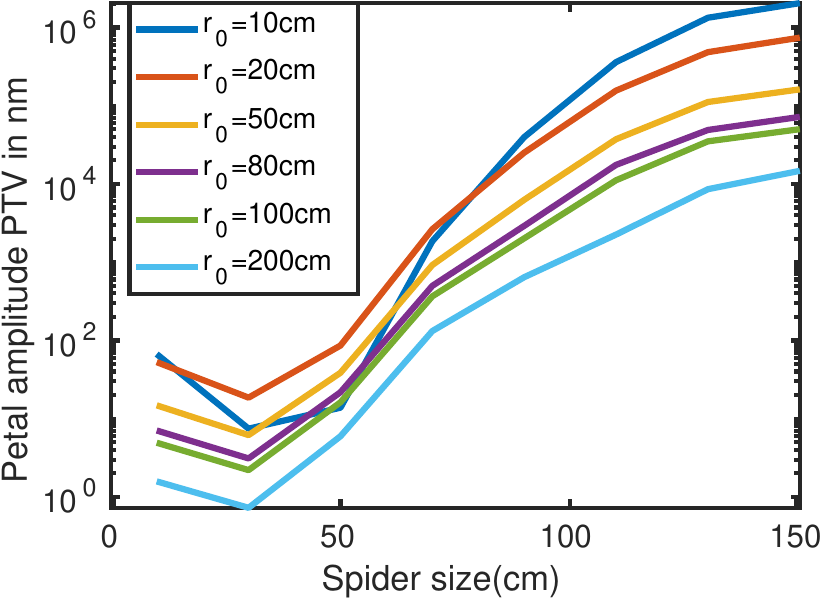}
        \caption{}
        \label{fig:petalvariation}
    \end{subfigure}
    \quad
     \begin{subfigure}[t]{0.45\columnwidth}
        %\centering
         \includegraphics[width=0.95\columnwidth]{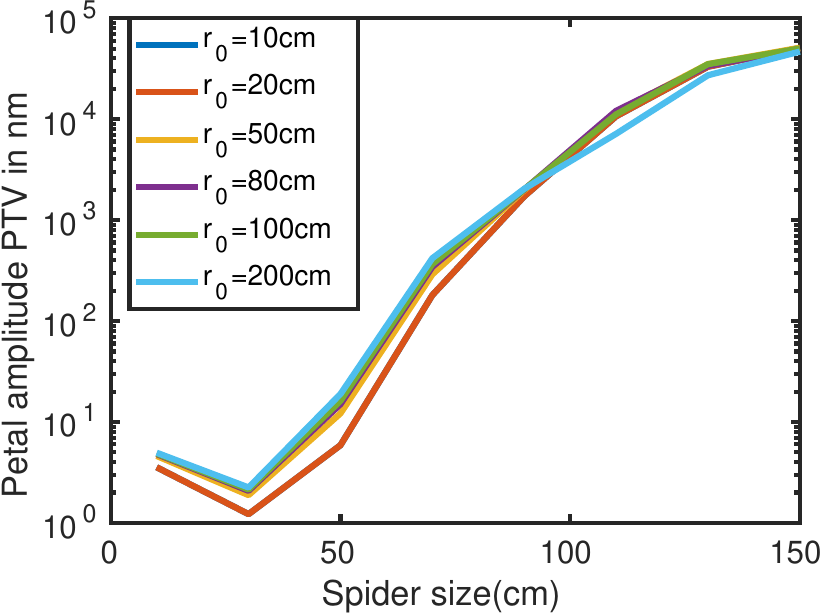}
         \caption{}
         \label{fig:petalvariation_correctedr0}
     \end{subfigure}
     \hfill
\caption{Petal mode with variable $r_0$ and spider size. Fig. a): Petal variation for large spiders and variable $r_0$ in first stage residuals. For spiders larger than the pitch of the ELT DM, the petal mode increases quickly, Fig. b) : Petal variation for large spiders and variable $r_0$ in first stage residuals corrected of atmospheric turbulence amplitude.}

\end{figure}

This adjustment reduces the differential piston to an atmospheric turbulence contribution under the spider rather than across the entire pupil. In this configuration, the turbulence is predominantly influenced by the size of the spider, and the $r_0$. In Fig. \ref{fig:petalvariation}, we simulate how the petal residuals of an AO loop using a modulated pyramid are impacted by the petal mode with minioning of the DM actuators (done in the simulation by using a continuous DM). In particular for the ELT-size spider at 50cm, with 15cm r0 (not plotted), we observe a mean petal residuals of 80nm Peak-To-Valley (PTV) in residual phases (= 40nm rms).
 When we correct by the amplitude change caused by $\phi(r_0)\propto \frac{1}{(r_0)^{5/3}}$, as shown in Fig. \ref{fig:petalvariation_correctedr0} we see that the amplitude of petal is directly proportional to the phase amplitude of the atmosphere. For the current SCAO instruments, this level of petal residuals falls within error budgets.

 There are multiple questions about the origin of the petal mode and ways to mitigate it. One possibility is to study the efficiency of different reconstructors. As this paper is tackling the petal problem by improving the sensor's measurement of petal mode, we used the simplest wavefront reconstruction available for our sensor: a matrix linear reconstruction using Intensity map, as presented in \cite{fauvarqueGeneralFormalismFourierbased2016}. It has been shown by \cite{bertrou-cantouValidationComposantsClefs2021} that an MMSE reconstructor exhibits similar performances to a minioned DM basis with a linear reconstructor, as reported in \cite{schwartzAnalysisMitigationPupil2018}. Notably, using a sensor that is not sensitive to the petal mode—such as the Shack-Hartmann in a centre of gravity measurement mode, or when the subapertures are smaller than the spider—can still significantly reduce the petal mode. However, the petal will achieve errors comparable to the remaining petal flares in a minioning system, as observed in \cite{bonnefondWavefrontReconstructionPupil2016}.

We detail in Sect. 2.2 the different sources of petal modes and why this approach is not suited for all of them. While effective for first-light instruments with purely atmospheric turbulence as the wavefront error, this method will not be adequate for all sources of petal mode. Additionally, it will not meet the error budget for second-generation instruments equipped with Extreme Adaptive Optics (EXAO).

This paper will aim to understand why the pyramid wavefront sensor cannot measure the petal mode and propose a second path in the AO system dedicated to petal mode sensing. We separate here the problem of the ELT phasing of the segments of the ELT (the $798$ x $1.45$m hexagonal mirrors which we don't consider) and the phasing of the fragments of the pupil (the 6 areas separated by the spiders). This would allow the measurement of residual petal mode after minioning and the measurement of exterior petal mode like Low Wind Effect (LWE) (explained in Sec. \ref{sec:petmode_source}), allowing EXAO instruments into the ELT. For the system to be quickly adaptable to an ELT instrument, we will use the same wavelength constraint as the HARMONI instrument with a sensing wavelength of $850nm$. As it must measure the fast-evolving atmospheric turbulence petal modes, it needs to be an AO-type sensor with a high sensitivity and linear reconstruction, so we will start with the already used AO sensor as the baseline for this paper.

\section{ State of the problem: the petal mode}

\subsection{Petal  Properties}

To reach the diffraction limit the petal mode would need to be lower than a few tens of nanometers in the residuals. 

 Uncorrected petal mode creates light residuals in the PSF comparable to slit interference patterns. The resolution on long exposure (with completely uncorrected petal mode) is then limited not by the size of the pupil of the telescope, but by the size of one fragment ($\approx $15m). This means petal mode can be responsible for a loss of resolution up to $ \sqrt{N_{\text{fragment}}}$. With the atmospheric petal only, the loss of resolution is not of this order, as the mode does not reach high values. However, LWE petal is expected to reach values over $\lambda$.
 From the differential piston between each fragment, we can define a petal mode basis constituted of $5$ orthogonal petal modes for the ELT (as appears in \cite{bertrou-cantouConfusionDifferentialPiston2022}). One particularity of the petal mode is that although it appears in the residual after an AO stage, it can be projected on modes that are measured and compensated by the AO stage. The most obvious is Tip-Tilt, onto which a lot of the ELT petal modes (or the one from our simplified pupil seen in Fig. \ref{fig:petalmodetoy}) can be projected. One might be tempted to orthogonalize this petal mode with the other mode of the basis. As the petal mode can be described in phase space as a Heaviside function, this can be done with an infinity of modes. The resulting petal, orthogonalized to a Zernike mode basis, will tend to a discontinuity phase mode around the spider. The orthogonalized petal mode resulting in this operation becomes very different from the mode appearing in AO residuals, such as \cite{schwartzAnalysisMitigationPupil2018}. To preserve the same mode as our studied mode, we choose to retain the original differential piston petal mode definition. However, to avoid confusion with other modes during phrase reconstruction, we also calibrate the basic Zernike modes (like tip-tilt, see Sec. \ref{sec:petalmode_rec} for more details).
To simplify the problem, we will consider a pupil with only one spider and, consequently, only one petal mode.

\subsection {Toymodel}

For a simplification of the problem, we will study a simplified pupil with a single spider and only 2 fragments and therefore only have 1 petal mode. This simple case scaled to a $10$m diameter telescope can be simulated with a coarser sampling and considerably reduces our computational time. To keep the representativity of the ELT case it uses the following parameters:  Spider size = $50$cm. $20$x$20$ DM (same pitch of a DM actuator = $50$cm as the ELT M4). 

\begin{figure}
    \centering
    \includegraphics[width=\columnwidth]{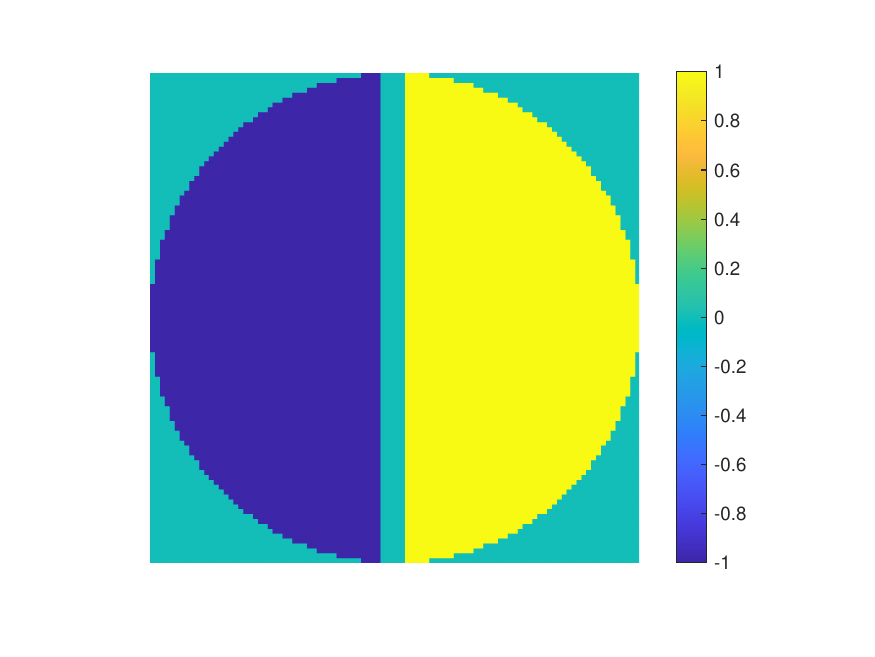}
    \caption{Toymodel petal mode}
    \label{fig:petalmodetoy}
\end{figure}

The petal mode is shown in Fig.\ref{fig:petalmodetoy}. We note that this petal is defined with an RMS amplitude of $1$ radian and thus has a PTV of $2$ radians between the left and right fragment. To stay consistent throughout this article, we will use an RMS unit for petal mode. With this normalization, an amplitude of $\pi$ radians RMS means a PTV of $2\pi$ radians, hence a $\lambda$ OPD wrap.

\subsection{Sources of petal mode} \label{sec:petmode_source}

 As was shown by \cite{bertrou-cantouValidationComposantsClefs2021}, the petal mode is badly sensed by the modulated PyWFS. % 
 This means an effect akin to the waffle mode can appear where the mode is amplified by the reconstruction of the AO. Furthermore, if the DM can create this mode, it will create loop instabilities. As the petal mode in this case comes from the AO loop and in particular its control, it evolves as fast as the AO correction.
 It is currently solved by a technique called minioning (\cite{bondHARMONIELTWavefront2022}). This technique acts on the control part of the AO loop by forbidding the creation of any differential piston by the DM. The remaining error is the atmospheric petal which exists under the spider. This residual atmospheric petal's amplitude depends on the size of the spider and the $r_0$ as was shown in the introduction in Fig. \ref{fig:petalvariation}.
 There are $2$ problems with this approach. First, this approach only works for atmospheric turbulence petalling. As the petal is not measured if it comes from other sources than atmospheric turbulence it will not fall within the acceptable first light instruments constraints. Then for the Extreme Adaptive Optic (EXAO) system, this is not an acceptable level of residual wavefront error. For a coronographic system for instance, while it is possible to design coronographs less impacted by this type of wavefront error( \cite{leboulleuxRedundantApodizedPupils2022}), it is at the cost of inner working angle. 

%\subsubsection{LWE petal}
Another source of petal mode in the phase is the LWE. It is a phenomenon that was discovered on the VLT during the first light of the SPHERE instrument and described in \cite{sauvageLowWindEffect2015}). This effect was mitigated on the VLT by improving the emissivity of the spider (i.e. the temperature of the spider is close to the temperature of the environment), but with the larger spiders of the ELT-class telescopes, LWE is expected to be stronger. A simulation of the airflow around the spiders has recently shown a $1\mu$m OPD around the spiders according to \cite{martinsTransientWavefrontError2022}, much larger than the petal residual after minioning. As the LWE is poorly understood we will consider it in this paper only as an uncontrolled source of petal mode. It will appear in the end-to-end simulation as a brutal petal mode appearing in the phase. It is to be separated from the atmospheric petal as it is a perturbation created by the telescope and not by the atmosphere though the final phase mode is equivalent.

\subsection{ Wavefront sensing measurement problem}

%\subsubsection{ lambda wrap}
The petal mode being a differential piston, means that in monochromatic light the whole phase screen wraps every $\lambda$. This makes a petal larger than  $\lambda$ impossible to detect correctly with a monochromatic sensor and thus the petal mode has a 'built-in' limited range of measurements. This comes back to a phase-unwrapping problem, that we leave aside for the study presented here. In this paper, we only consider monochromatic light and test whether a measurement can be done accurately in ELT conditions.

\subsection{Solutions proposed to the petal problem in the literature}

Any slope sensor will not be able to measure the petal mode in a subpupil. It is not a problem when the spiders are small like on the VLT where the petal created by the spiders is negligible. But with 50cm spiders, it becomes a mode large enough to decrease EXAO performances.
 Phase sensors like the Zernike Wavefront Sensor (ZWFS) or the unmodulated PyWFS (uPyWFS) have been proposed as their intrinsic response is more sensitive to phase discontinuities. However, their dynamic is not large enough to measure the atmosphere at the same time.

There are two ways to solve the petal problem proposed in the literature: modify the AO wavefront sensor, or add an additional sensor dedicated to measuring the petal mode, a petalometer. 
The first solution appears in the METIS instrument. This instrument senses the Wavefront at a longer wavelength. Thanks to a lower turbulence phase and more linear sensor, it doesn't show any petal mode in its residuals (see  \cite{hipplerSingleConjugateAdaptive2019}  and \cite{carlomagnoMETISHighcontrastImaging2020}). Another solution is the flip-flop method proposed by \cite{englerFlipflopModulationMethod2022a} where the modulation of the PyWFS is cut to use a temporary uPyWFS and measure the petal mode in the residual.

The GMT has opted for the second solution with the development of the Holographic Dispersed Fringe Sensor (HDFS, see \cite{haffertHolographicDispersedFringe2022}), a sensor dedicated to the phasing of the GMT mirrors. 

 Moving all AO systems to IR looks like a simple solution but there are a lot of advantages to keeping the sensing in the visible light. IR detectors are slower and noisier than their visible counterparts, meaning fainter stars can be used as NGS by visible systems. Astronomical observations are mainly using IR for most of the first-light instruments. We make our simulation with the wavelength used in the HARMONI instrument: $850$nm. We instead study the petalometer approach in the rest of this paper.

\section{Measurement of petal mode in AO residuals}

\subsection{2-Path AO system}
\begin{figure}
\centering
\includegraphics[width=\columnwidth]{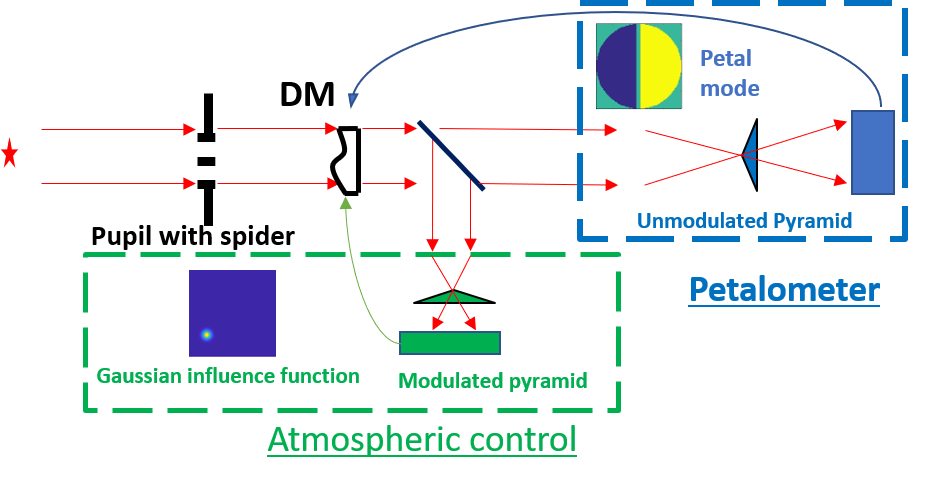}
\caption{2 Path sensor scheme considered. The green path is dedicated to atmosphere wavefront sensing and features a modulated PyWFS and a classical $20$x$20$ actuator DM. The blue path is dedicated to the sensing of petal mode and features an uPyWFS and controls only the petal mode of the DM}
\label{fig:2path_sensor}
\end{figure}

As was shown by \cite{bertrou-cantouValidationComposantsClefs2021} the uPyWFS produces a strong signal for the petal mode even in the presence of spider compared to the modulated PyWFS. However, it is not able to measure the petal mode accurately among the whole atmospheric turbulence for a realistic $r_0$. It is proposed instead to use the uPyWFS to measure the petal mode among the residuals of an AO loop, as a petalometer. The question in this part is whether the uPyWFS can reconstruct accurately the petal mode among AO residuals. To test this capability we analyze the petal mode reconstructed by the PyWFS used as a petallometer with the 2path system shown in Fig. \ref{fig:2path_sensor}.

There is a single Deformable Mirror (DM) correcting the aberrations in the AO loop. This mirror is described by a modal basis including a pure petal mode of our toymodel pupil, as well as Gaussian influence functions of a $20$x$20$ regular actuator grid. The Gaussian influence functions are controlled by the modulated PyWFS and are dedicated to the atmospheric turbulence compensation. The second path includes a petalometer which controls the pure petal mode of the DM. Diverse sensors could be proposed as petalometers for the second path. As we know that the uPyWFS is a possible alternative in terms of sensing, we use it in this first test. The uPyWFS only controls the petal mode though it can measure other modes.

The first source of petal mode is atmospheric turbulence. We use this configuration to take advantage of the reduced atmospheric petal mode in the residuals after a first stage using a minioning DM. The petalometer aims to measure and allow the control of both atmospheric petals and LWE. We also added a fixed petal after some iterations to simulate LWE.

\subsection{uPyWFS petal mode reconstruction} \label{sec:petalmode_rec}
We use the Intensity Map method described in \cite{fauvarqueOptimisationAnalyseursFront2017} so using all the pixels. 
The linearity curves presented later have been reproduced with the slopes map method and have given the same results. There does not seem to be a preferred method for measuring petal mode. 
The first step is to calibrate the interaction matrix of our system. We create this interaction matrix by simulating the Intensity map for a variety of modes.
 For the atmospheric control, the calibrated phase modes will be the zonal base of the 400 actuators. For the petal mode interaction matrix, we need not only to measure the petal mode but also other Zernike modes in particular tip and tilt to avoid confusion between these low-order modes and the petal modes. In practice, we calibrate a modal basis containing the Petal mode + 30 Zernikes phase modes.
\begin{equation}
        \delta I(\phi)= \frac{I(\epsilon \phi)-I(-\epsilon \phi)}{2*\epsilon}
        \label{eq:delta-intensity}
    \end{equation}
    with I($\phi$) the intensity on the PyWFS detector for a given phase $\phi$
     $\epsilon$ a factor small enough to work in the linear regime of the PyWFS. In simulation, we use $10^-10$
    $\delta I(\phi)$ is the push-pull intensity map of the mode $\phi$

\begin{equation}
    \mathcal{D} = (\delta I(\phi_{1}),..., \delta I(\phi_{i}),..., \delta I(\phi_{N}))
\end{equation}
See equation \ref{eq:delta-intensity} for the computation of $\delta I(\phi_{1})$
Then the interaction matrix is inverted using a Moore–Penrose pseudo inverse to get the control matrix \(D^\dagger\). We use a conditioning number of 100. Assuming small phase we should have the relationship : 
\begin{equation}
\hat{\phi}= \mathcal{D}^{\dag}\Delta I(\phi)\\
\end{equation}

We express any phase as its decomposition on the modal basis used 
\begin{equation}
\hat{\phi} = \sum_{i=1}^N a_i\phi_i \\
\end{equation}
with \(a_i\) the amplitude of the mode \(\phi_i\)
\begin{equation}
\begin{pmatrix}
\hat{a_1} \\
\hat{a_2} \\
...\\
\hat{a_N}
\end{pmatrix}
= \mathcal{D}^{\dag}\Delta I(\phi)\\
\end{equation}
with \( \hat{a_i}\) the estimated amplitude of the mode \(\phi_i\).
%\subsubsection{linearity curve}
\subsection{Linearity curve of petal mode reconstruction}
We want to test the measurement of petal mode with a PyWFS used as a petalometer. To that end, we compute the linearity curve to a petal mode first without AO residuals to set the ideal case, and then with typical AO residuals. To plot this linearity curve we take a given phase screen and add a given petal mode amplitude, varying between $[-\pi : \pi]$. As we are using monochromatic light the signal is wrapped outside of these boundaries and computing the linearity curve for higher amplitude is of no use.

%\paragraph{ linearity curve in absence of residuals}

In the absence of residuals we see the expected result previously shown by \cite{espositoCophasingSegmentedMirrors2003}:
\begin{equation}
\hat{a_1}=sin(a_1)  
\end{equation}
with $\hat{a_1}$ the estimated petal mode amplitude for an input of $a_1$ expressed in rms value. 
This expression emphasizes the wrapping of the petal mode estimation with monochromatic light. In the absence of residuals, there is no difference in whether there is a spider or modulation as these are noiseless tests. 
%Despite the linearity of the reconstruction method, 
We see here the intrinsic non-linearity of the PyWFS and specifically the non-linearity of the petal mode itself. The linearity of both modulated and uPyWFS is tested here because the first question that arises with any feature is whether the feature would remain with a modulated pyramid. However, it is inefficient for a real system to try to reconstruct petal mode with a modulated PyWFS and we only keep this curve to demonstrate that the problem remains.

\subsection{linearity curve in the presence of residuals}

\begin{center}
    \begin{tabularx}{\columnwidth}{| >{\centering\arraybackslash}X | >{\centering\arraybackslash}X |}
        \hline
        Pupil & Diameter = $10$m \\& Monolithic primary mirror\\ 
        & Variable spiders (reference case at $50$cm) \\
        \hline
        Turbulence & $3$ Layer $r_0$=$15$cm@$550$nm\\& wind speed=$5$m/s \\& Outer scale $L_0$=30m\\
        \hline
        DM & $50$cm pitch ($20$x$20$), square pattern, gaussian influence functions, 0.3coupling\\ 
        \hline
        PyWFS & $100$x$100$ subapertures\\& no noise (photon or readout)\\ &Atmospheric control modulation = $3\lambda/D$ \\
        \hline
        Target & $\lambda=850nm$ \\& on-axis star \\
        \hline
        Controller & Loop Rate 1000Hz\\& 1 Frame delay (+integration) \\& Matrix-Vector-Multiplication + Integrator
        \\
        \hline  
\end{tabularx}
\begin{table}[!h]

\caption{Toymodel simulation parameter}
\label{table:AO_simconditions}
\end{table}

\end{center}

We simulate the AO residuals with an AO loop using the parameters summed up in table \ref{table:AO_simconditions}.  As the petal mode has very high frequency parts aliasing could be the source of a lot of issues. To reduce that problem we will be using a large oversampling here with 100x100 subapertures. All the further results have been confirmed with a 20x20 subapertures case. The resulting residuals have an amplitude of 120nm RMS and give a 55\% SR at the top of the pyramid. The residuals are simulated without a spider and filtered from petal mode to make sure that the further injected petal will be the only petal mode present in the phase. The tip-tilt which could be changed by the petal mode filtering has a negligible amplitude as it is already phase residuals and has an RMS amplitude of 14nm.

\begin{figure}
 \begin{subfigure}[c]{0.3\columnwidth}
 %\savebox{\tempfig}{\includegraphics[width=0.5\columnwidth]{PDF_version_of_plots/fullpupil_linearity_test.pdf}}% Store larger image in box
    %\raisebox{\dimexpr\ht\tempfig-\height}{ 
    \includegraphics[width=\columnwidth]{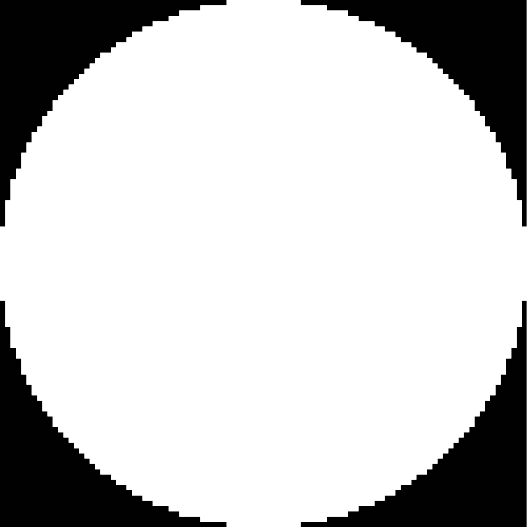}
 \caption{}
 \end{subfigure}
 \hfill
 \begin{subfigure}[c]{0.65\columnwidth}
 \centering
\includegraphics[width=\columnwidth]{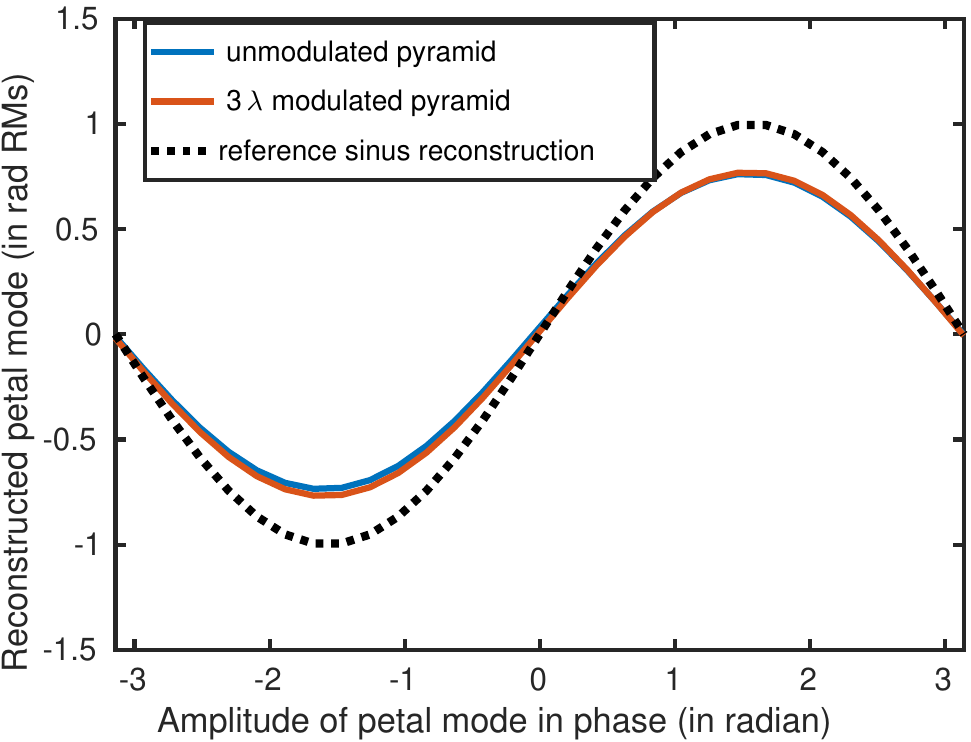}
\caption{}
\label{fig:linplotRESfullpupil}
\end{subfigure}
\begin{subfigure}[c]{0.3\columnwidth}
 \centering
  %\savebox{\tempfig}{\includegraphics[width=0.5\columnwidth]{PDF_version_of_plots/fullpupil_linearity_test.pdf}}% Store larger image in box
   % \raisebox{\dimexpr\ht\tempfig-\height}{  
   \includegraphics[width=\columnwidth]{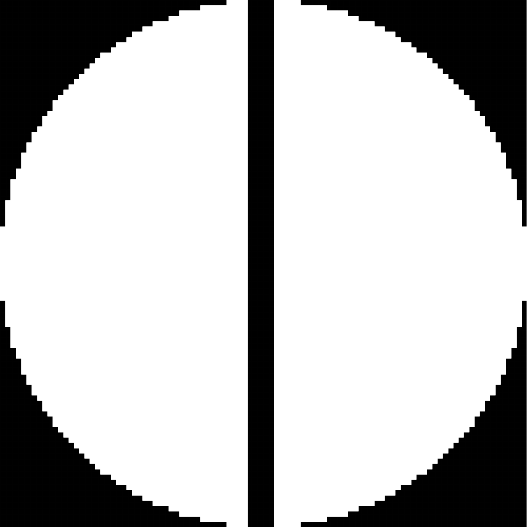}
 \caption{}
 \end{subfigure}
 \hfill
 \begin{subfigure}[c]{0.65\columnwidth}
 \centering
\includegraphics[width=\columnwidth]{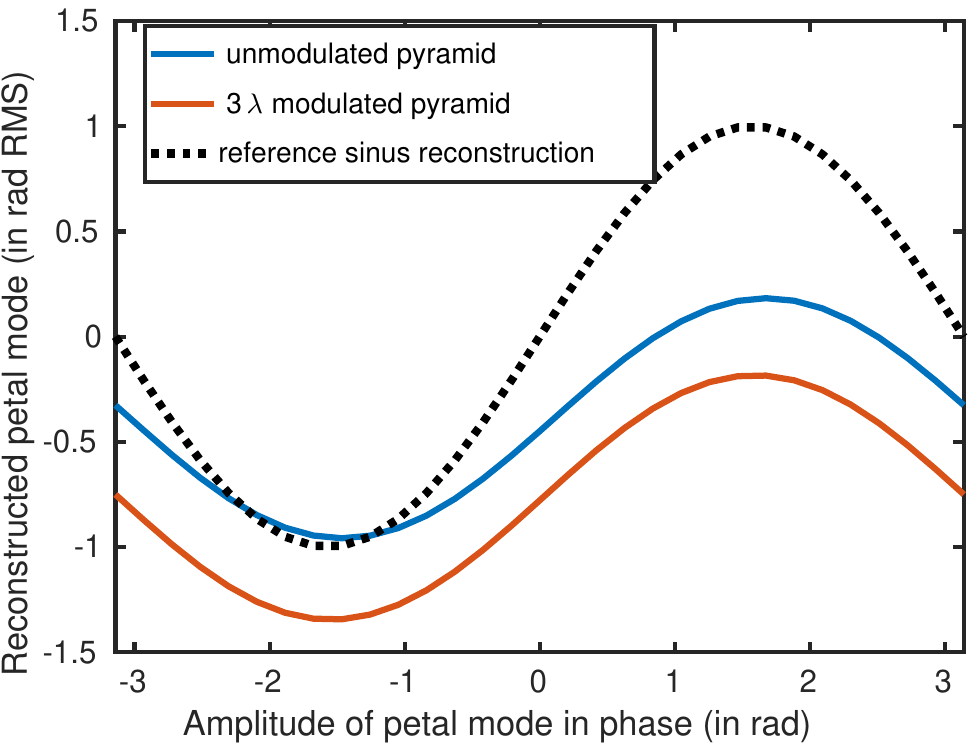}
\caption{}
\label{fig:linplotREStoymodel}
\end{subfigure}
\hfill
\caption{Linearity curve in presence of residuals for unobstructed pupil and toymodel pupil. Fig. a) : Unobstructed pupil. Fig. b) : Linearity curve of petal mode around residual phase screen for unobstructed pupil. Fig. c): Obstructed pupil. Fig.d) : Linearity curve of petal mode around residual phase screen for toymodel pupil}
\label{fig:linplot_withresidual}
\end{figure}

\begin{figure}
\centering
\includegraphics[width=\columnwidth]{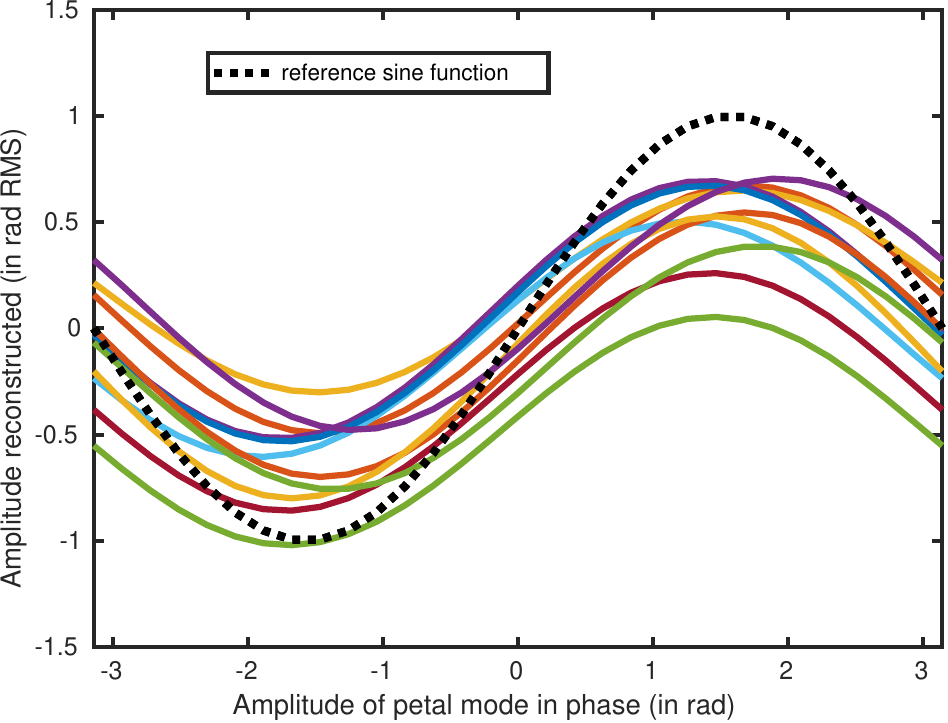}
\caption{Linearity curve with $10$ independent residuals with uPyWFS}
\label{fig:indepresidualslinplot}
\end{figure}

The resulting linearity curves are very different between the unobstructed pupil and toymodel case as seen in Fig. \ref{fig:linplot_withresidual}. In the unobstructed pupil case, we see mostly the Optical Gains (OG) (Fig. \ref{fig:linplotRESfullpupil}) $g$ which reduce the amplitude of the reconstructed petal mode. We can make a separation here between the optical gains coming from all the other modes present in the residuals $g$, and the optical gain coming from the petal mode itself, the $\sin$ function.%\jeff{On peut discuter de ça. Je pense pas que le sinus soit un "gain optique", c'est plutôt la réponse normale de la pyramide à un mode pétale. La réponse théorique c'est une sinusoide parceque la pyramide est intrinsèquement un senseur interférométrique non ?}
\begin{equation}
\hat{a_1}=g*sin(a_1). 
\end{equation}
With the spider present, the linearity curves are offset. It means that the uPyWFS measures a petal mode amplitude when there is no petal mode in the phase, it confuses another mode with petal mode. This confused mode is seen in the linearity curve as a fixed offset value added to the sinusoidal that we note $c$ and call the "Petal Confusion". Furthermore, as is shown in Fig. \ref{fig:indepresidualslinplot}, $c$ depends on the AO residual.
\begin{equation}
\hat{a_1}=g*sin(a_1)+c. 
\end{equation}
The typical amplitude of petal confusion $c$ is a few tenths of a radian, which makes the value reconstructed by the PyWFS for a null input very far from the true value of petal ($0$ here). We need to understand this parameter to use the petalometer efficiently.
$c$ is computed by taking the mean of the linearity curve. There is also sometimes a term of dephasing appearing. The first possible origin of this dephasing would be a petal mode in the residuals, but in this simulation, we made sure to specifically filter it. We consider it as another defect of the reconstruction. Examples can be seen with various residual on Fig. \ref{fig:indepresidualslinplot}
\begin{equation}
\hat{a_1}=g*sin(a_1+d)+c. 
\end{equation}

We also note that for some extreme cases the petal confusion is so large that the linearity curves only estimates non-zero values of petal mode.
 Its phase reconstruction doesn't cross the $0$ petal mode measured line. 
Another important parameter is its speed. each curve plotted in Fig. \ref{fig:indepresidualslinplot} is separated by $500ms$. The petal confusion changes fast with each phase screen.

\subsection{Dependence of petal confusion to AO residuals}

\begin{figure}
\centering
\includegraphics[width=\columnwidth]{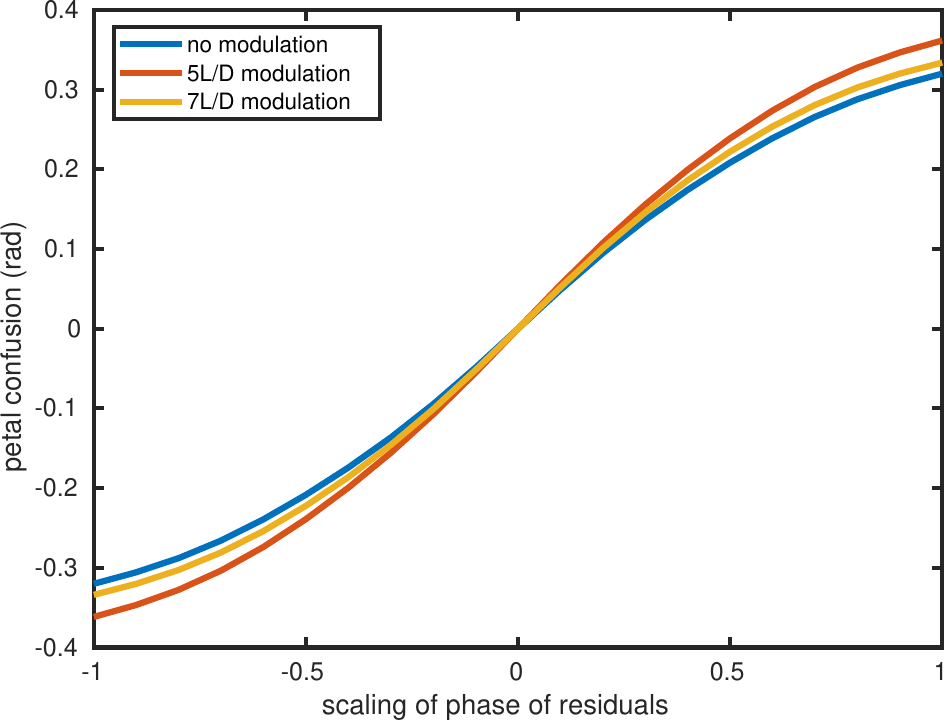}
\caption{Petal confusion variation with change of phase residual amplitude}
\label{fig:Scalar_test-multmod}
\end{figure}

The first possible origin of this petal confusion comes from the non-linearity of the uPyWFS.
We test the dependence of the petal confusion on the amplitude of AO residuals. To do so, we plot the petal confusion (as the average value of the linearity curve) with respect to the amplitude of AO residuals. The scaling parameter is a multiplier on the original AO phase residuals. 
To this end, we use the same residual phase screen as before and scale them by a multiplicative factor \(s\in[-1 , 1]\). The dependence curve of $c$ with respect to AO residuals is plotted in Fig. \ref{fig:Scalar_test-multmod} as a function of this scalar. $10$ uncorrelated residual realizations were used and the curves of amplitude with scaling were averaged.
 When $s$ is close to $0$ we can write : $c(s*\phi)=s\times c(\phi)$. The petal confusion is very much a linear effect.

\subsection{Origin of the petal confusion}
The origin of the petal confusion seems nonetheless to be in the phase residuals. We have found a specific shape it takes for the PyWFS, but its effects were already reported for the PyWFS and Zernike wavefront sensor by \cite{bertrou-cantouConfusionDifferentialPiston2022}. Furthermore, it seems to be a linear effect as was shown by the scalar test. The linearity of the confusion effect means we can construct the phase mode associated with it. 

\begin{figure}
\centering
\includegraphics[width=\columnwidth]{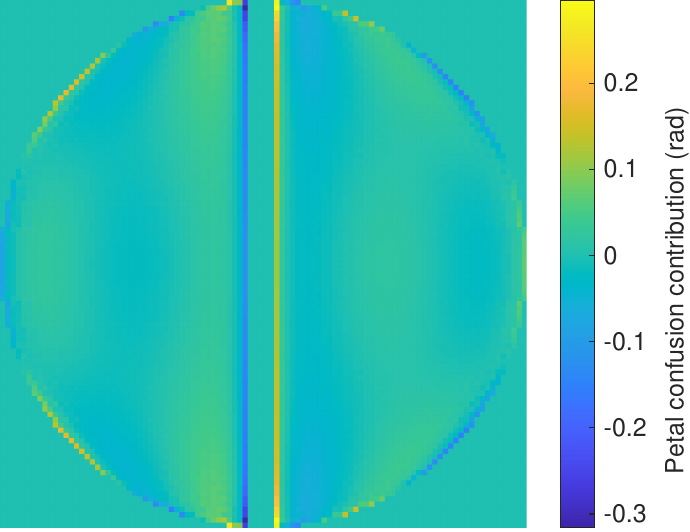}
\caption{Petal confusion map}
\label{fig:petalconfmap}
\end{figure}

To that end, we compute a map of the confusion, e.g. how much each phase pixel creates petal confusion separately. As it is a linear effect we can then sum the petal confusion contribution of each pixel into a phase map.
 For this construction, we put all phase pixels at $0$ except one which is at 1 rad. We orthogonalize this phase pixel to the petal mode to avoid our reconstruction creating petal mode.  We use this phase as the residual and test the linearity of the petal mode with this residual.
 Then we compute the petal confusion caused by this pixel \(c_{x,y}\). We then plot the full map of confusion shown in Fig. \ref{fig:petalconfmap}.

This map, which we refer to as "confusion map", is expressed in phase space and can be interpreted as the phase mode responsible for the petal confusion. 
The mode creating petal confusion seems to have two distinct parts: a high-frequency line on each side of the spider (or discontinuity mode), and a low spatial frequency sinusoidal phase.
 This low-order mode comes from the number of modes calibrated in the interaction matrix for the reconstruction of the petal mode. This stems from the petal mode which can be projected on an infinity of spatial frequencies. The high-frequency signal always appears even with very high-frequency Zernike integrated in the petalometer interaction matrix. This mode is too high frequency to be controlled by our DM and cannot be separated from the rest of the typical AO residuals. As petal mode contains high frequency due to its discontinuity talking about a high frequency being reconstructed as a lower frequency is tricky but petal confusion (in particular the spider discontinuity component) can be seen as a form of aliasing. It is not a classical aliasing as oversampling the phase measurement was tested and doesn't solve the problem.

The conclusion of the analysis on petal confusion identification is there will always be some petal confusion that cannot be separated from residuals. Therefore to measure petal mode efficiently we need to reduce the effect of AO residuals to increase the accuracy of the petal mode measurement.

\section{Reduce the impact of residuals on petal mode measurement }

To reduce the impact of the residuals on the signal, a solution would be a longer integration time, or moving to a longer wavelength for sensing.
 This is not compatible with our original aim which was a fast measurement in the visible, so we need a new strategy to reduce the impact of residuals on the measurement.

\subsection{Reduce the effect of residuals in phase space}

\begin{figure}
\centering
\includegraphics[width=\columnwidth]{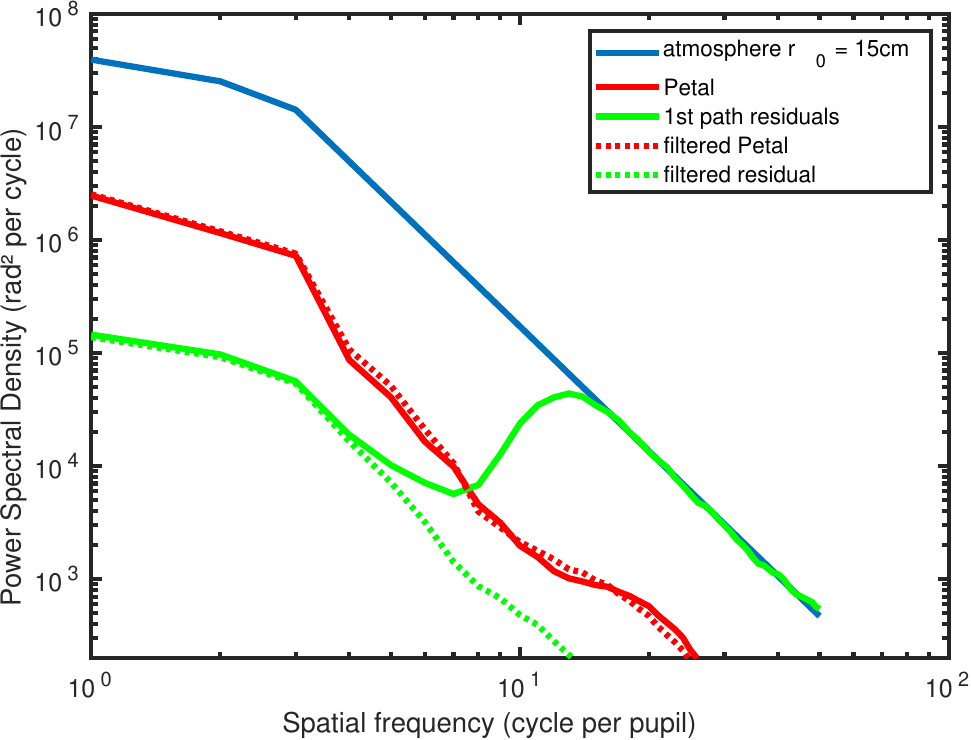}
\caption{PSD comparison Atmosphere and first path Residual and petal mode without (plain line) and with 5$\lambda$/D filter (dotted line)}
\label{fig:PSDatmRESpetalfiltersize5}
\end{figure}

%\subsubsection{PSD of petal mode and residuals} 
We now compute the PSD of atmosphere contribution and the PSD of the petal mode. The residuals PSD is computed using the same first stage system as in part 4 and then their PSF are averaged over $1000$ independent residual phase screens. We can see in Fig.\ref{fig:PSDatmRESpetalfiltersize5} (solid lines) that they have a different distribution in the spatial frequency domain. The PSD of petal mode is plotted for the same RMS amplitude of petal mode as the RMS amplitude of residuals (120nm RMS).% \jeff{value of petal mode for this comparison ?} 
  Residuals have lower PSD values than petal mode in the low spatial frequencies but dominate in the high spatial frequencies. So if one can filter selectively to keep only the low spatial frequencies, it makes the separation between the petal mode and residuals easier. One can consider the residual as a form of noise on our petal mode measurement, then we want to improve the SNR by filtering selectively the residuals. In the focal plane, there is already this organization by frequency and there is a focal plane already accessible when using a PyWFS: the tip of the PyWFS. We have a new kind of sensor adapted to be a petalometer: a spatially filtered uPyWFS (SF+uPyWFS).

It is to be noted that \cite{usudaPreliminaryDesignStudy2014} already proposed such a PyWFS for their second WFS channel to lift the $\lambda$ uncertainty but with a reverse filtering. It was proposed for GMT to filter the low-order frequencies with a chip hiding the heart of the PSF. When looking at the PSD, the petal mode does indeed evolve differently at higher frequencies than the atmosphere or residuals.  Due to the discontinuity in the phase, at high spatial frequencies, the petal mode PSD is over the atmospheric PSD for a comparable amplitude. In terms of petal confusion that would mean a reduction of the low-order part of the petal confusion so it would be interesting to test it in a further paper.

%\subsubsection{effect of size of spatial filter on phase}
We simulate the effect of a spatial filter at the tip of the pyramid, as observed from the pupil phase. In this example, we start with the electric field in the entrance pupil plane, propagate it to a focal plane, use a focal plane filter (a circular pinhole), and propagate it to a pupil plane following Fig. \ref{fig:phasefilteredscheme}. 

 We then computed the PSD of residuals and petal mode after focal plane filtering (Fig. \ref{fig:PSDatmRESpetalfiltersize5}). The pinhole size used for this computation is $5\lambda/D$ of radius.

We see that the first path residuals are suppressed before the petal mode by this focal plane filter. The balance has to be found between reducing the first stage residual and keeping the spatial filter opened to let as much light as possible still. 
To decide what would be the best size of filter we can think in terms of Signal to Noise Ratio. The Signal we are interested in is the petal mode. The 'noisy signal' are the residuals.
 We compute the variance of residuals and variance of petal mode with different Spatial Filter size (SF size). 
\begin{equation}
SNR(SF size)=\frac{\sigma^2_{petal|filtered}}{\sigma^2_{residual|filtered}}
\end{equation}
With our simulation parameter, the $2\lambda/D$ spatial filter has the best phase SNR (see Fig. \ref{fig:placeholder_SNRspatialfitersize}). However, having a small SF means limiting the intensity entering the WFS. There is a compromise to find here between SNR and loss of intensity. We make the following simulation with a 5$\lambda/D$ spatial filter. With this spatial filter, we have a SNR of $4$ and lose $50\%$ of the intensity.
 In Fig. \ref{fig:placeholder_SNRspatialfitersize} curve we see a drop of SNR starting at $10\lambda/D$. this is due to the residual of the AO (in particular the fitting error) which appears as the intensity at a spatial frequency higher than $10\lambda/D$. In practice, the spatial filter doesn't filter the AO residuals anymore if it is larger than this radius.

\begin{figure}

\begin{subfigure}[t]{0.8\columnwidth}
 \centering
 \includegraphics[width=\columnwidth]{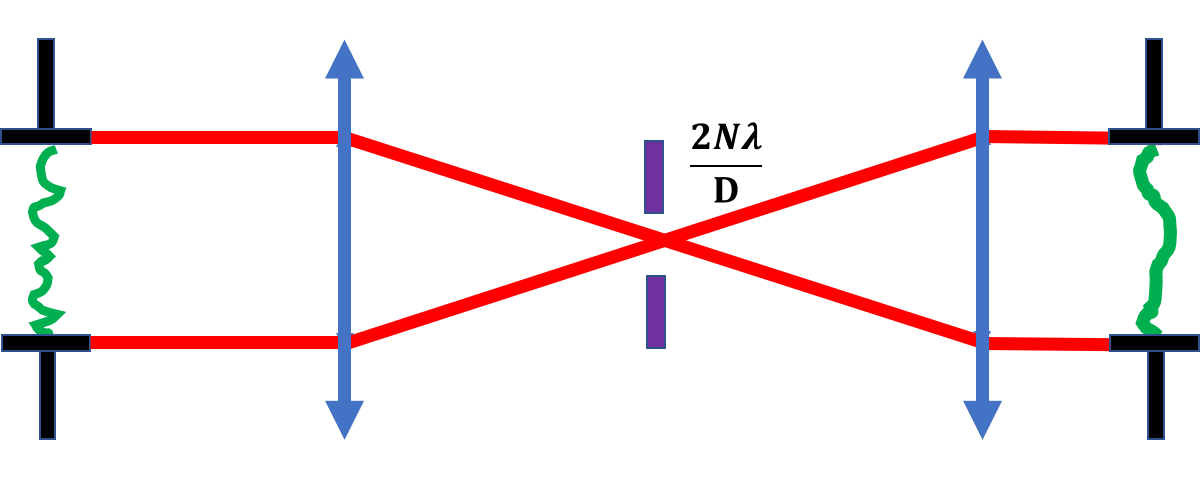}
 \caption{}
 \label{fig:phasefilteredscheme}
 \end{subfigure}
 \quad
\begin{subfigure}[t]{0.8\columnwidth}
 \centering
 \includegraphics[width=\columnwidth]{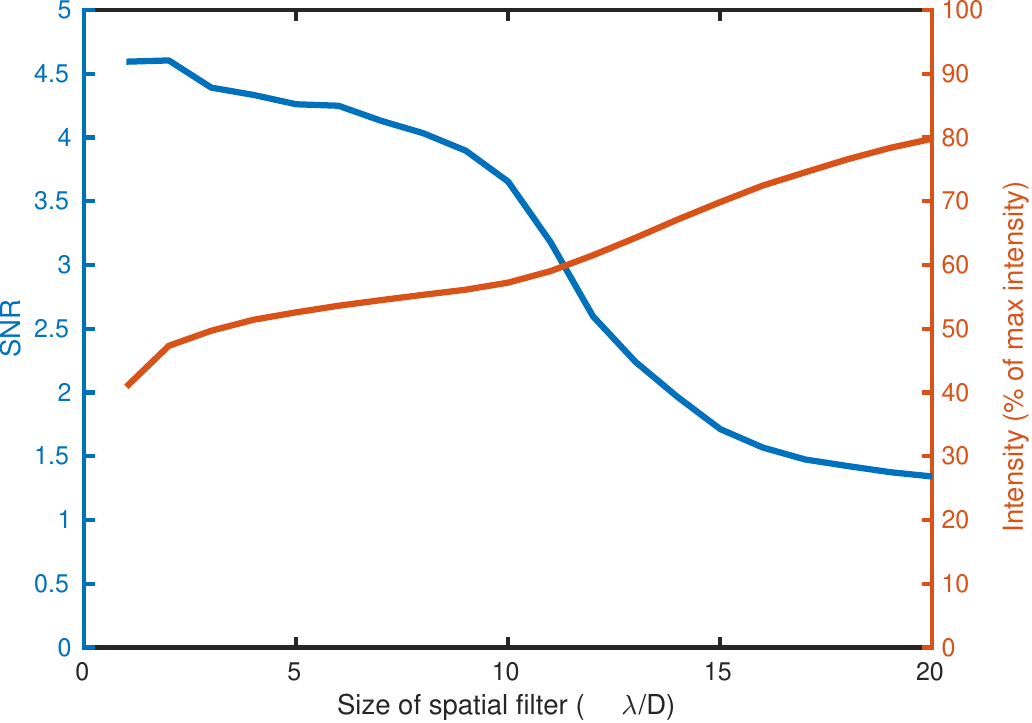}
 \caption{}
 \end{subfigure}
 
 \hfill
    \caption{Petal mode to Residual ratio for variable spatial filter size. Fig. a): Optical path considered for the spatial filtering tests , Fig. b) : Signal to noise ratio for petal mode and first path residuals compared to the intensity going through the spatial filter}
    \label{fig:placeholder_SNRspatialfitersize}
\end{figure}

\subsection{effect of the spatial filter on PyWFS signal}

%\subsubsection{effect on delta intensity}
A spatially filtered PyWFS (SF+PyWFS) is implemented in simulation. From a mathematical point of view, a SF+uPyWFS has a phase (PyWFS) and amplitude mask (SF) in the focal plane. A modulated PyWFS needs the spatial filtering step before the modulation step and is not simply a change of the focal plane mask.
 
\subsection{Effect of spatial filter on linearity curve of PyWFS}
\begin{figure}
    \centering
    \includegraphics[width=\columnwidth]{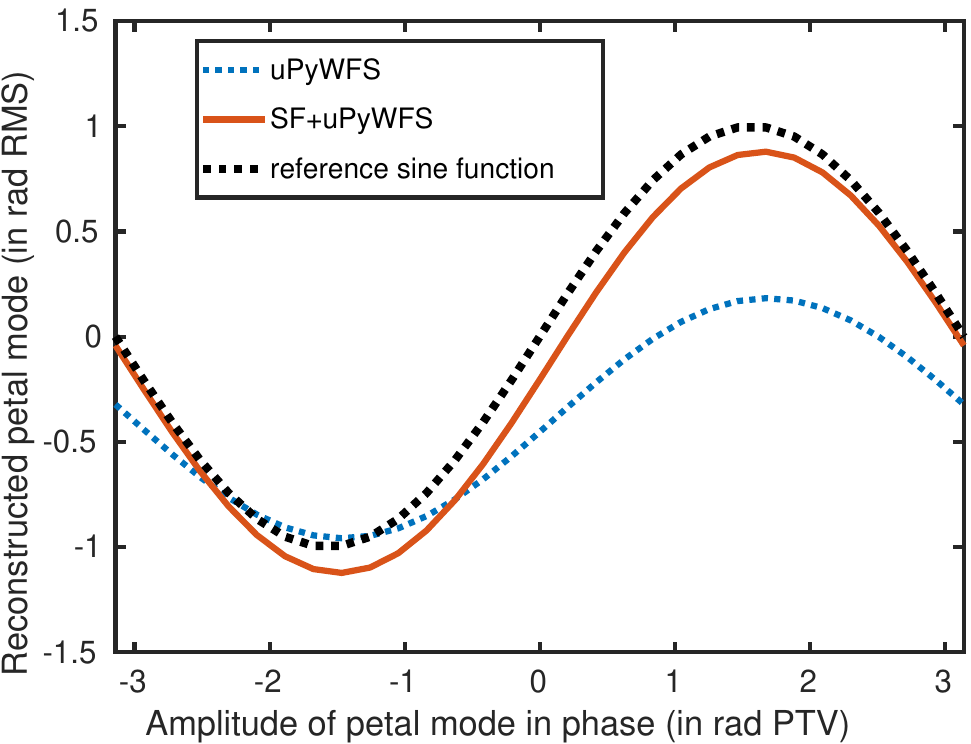}
    \caption{Comparison of the petal mode reconstruction between a uPyWFS and SF+uPyWFS}
    \label{fig:spatial_filter_linearity}
\end{figure}

The previous linearity test was done using the SF+uPyWFS as the second path petalometer (Fig. \ref{fig:spatial_filter_linearity})

\begin{figure}
    \centering

            \includegraphics[width=0.8\columnwidth]{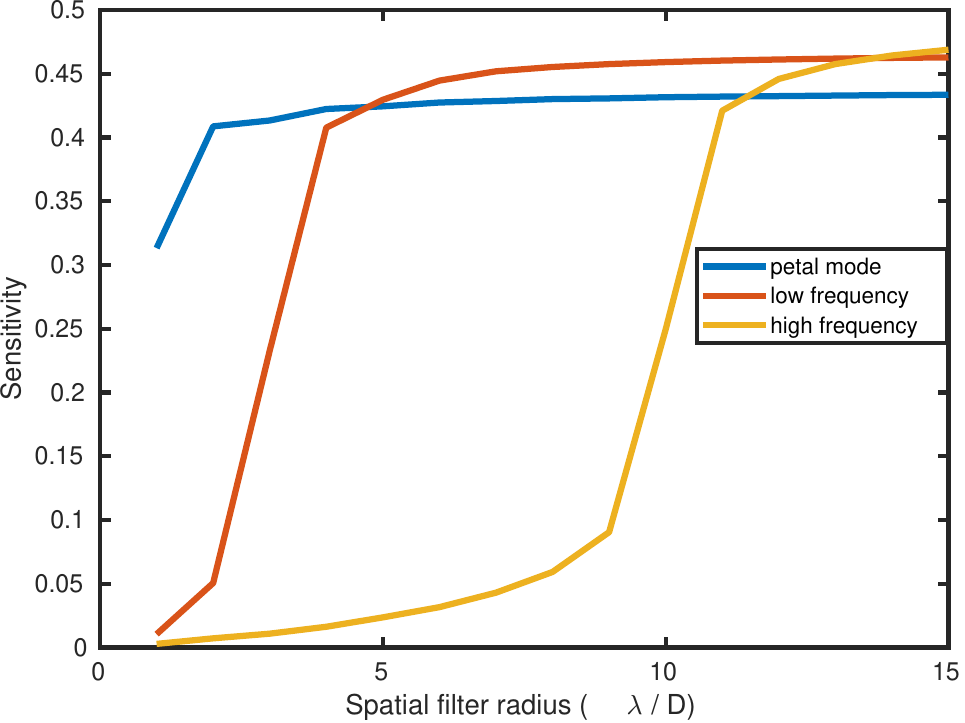}
    \caption{Comparison of sensitivity of 3 modes for various spatial filter radius. Low spatial frequency is a $3\lambda /D$ sinus and high spatial frequency is $10\lambda /D$ sinus}
    \label{fig:sensitivitywithSFsize}
\end{figure}

\begin{figure}
\centering
            \includegraphics[width=0.8\columnwidth]{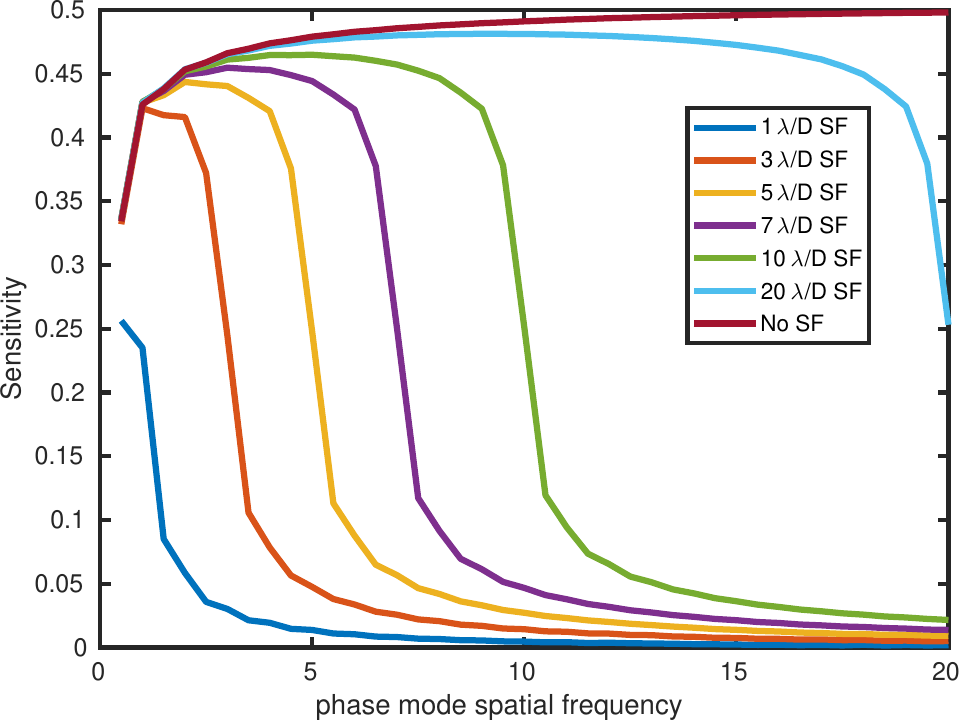}
    \caption{ Comparison of sensitivity to pure spatial frequency modes. We see that the SF+uPyWFS has the same sensitivity as the uPyWFS until the spatial frequency are close to the SF radius.}
    \label{fig:sSensitivitycomparison}
\end{figure}

\begin{figure}
    \centering
    \begin{subfigure}[t]{0.8\columnwidth}
    \includegraphics[width=\columnwidth]{PDF_version_of_plots/multiple_residual_linearity_test.pdf}
        \caption{}
    \end{subfigure}
    \quad
    \begin{subfigure}[t]{0.8\columnwidth}
    \includegraphics[width=\columnwidth]{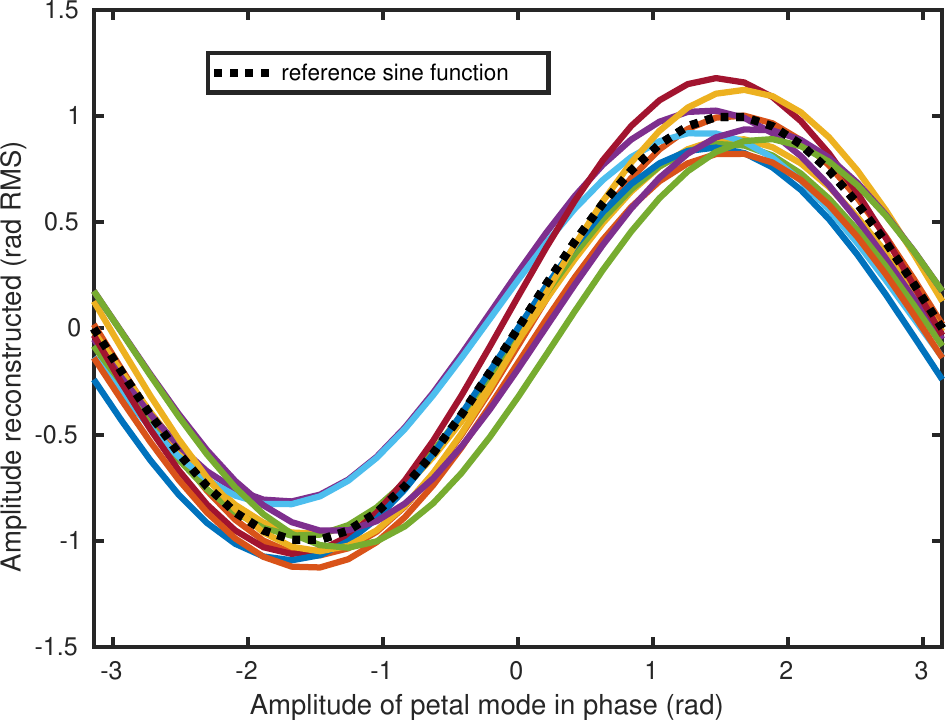}
    \caption{}
    \end{subfigure}
    \caption{ Petal mode reconstruction with 10 independent 1st path residuals. Comparison between uPyWFS and SF+uPyWFS behavior. The optical gain are greatly reduced as well as the petal confusion (reduced by a mean of a factor 6. Fig.a) : Petal mode reconstruction with 10 independent residuals with uPyWFS presented earlier (Fig. \ref{fig:indepresidualslinplot}). Fig. b): Petal mode reconstruction with $10$ independent residuals with SF+uPyWFS}
    \label{fig:multiple_linpot_SF}
\end{figure}

There are two remarkable effects of spatial filtering. The first is less impact on the optical gains. As the spatial filter reduces the residuals drastically, the PyWFS is used in a regime closer to the low aberration approximation. Hence the linearity of the signal improves visible here as OG are closer to 1. 
Another effect is in the modes impacted by the spatial filter. The sensitivity to modes now drops when the spatial filter radius is under the spatial frequency of the mode. 
In Fig.\ref{fig:sensitivitywithSFsize} we compare the sensitivity of petal mode, a $3\lambda /D$ and $10\lambda /D$ sinusoïdal mode with respect to spatial filter size. The sensitivity drops fast once the modulation radius is under the spatial frequency of the phase mode. In Fig.  \ref{fig:sSensitivitycomparison} we computed the sensitivity of different SF+uPyWFS compared to uPyWFS (dark red curve). 
These two effects make the measurement of petal mode with a SF+uPyWFS more accurate, as is visible with multiple different residuals in Fig. \ref{fig:multiple_linpot_SF}. Compared to Fig. \ref{fig:indepresidualslinplot}, the petal mode reconstruction is closer to the expected sinus reconstruction when using a SF+uPyWFS.

\section{Performance on an AO system assisted by a petallometer}

\begin{figure}
    \centering
    \includegraphics[width=\columnwidth]{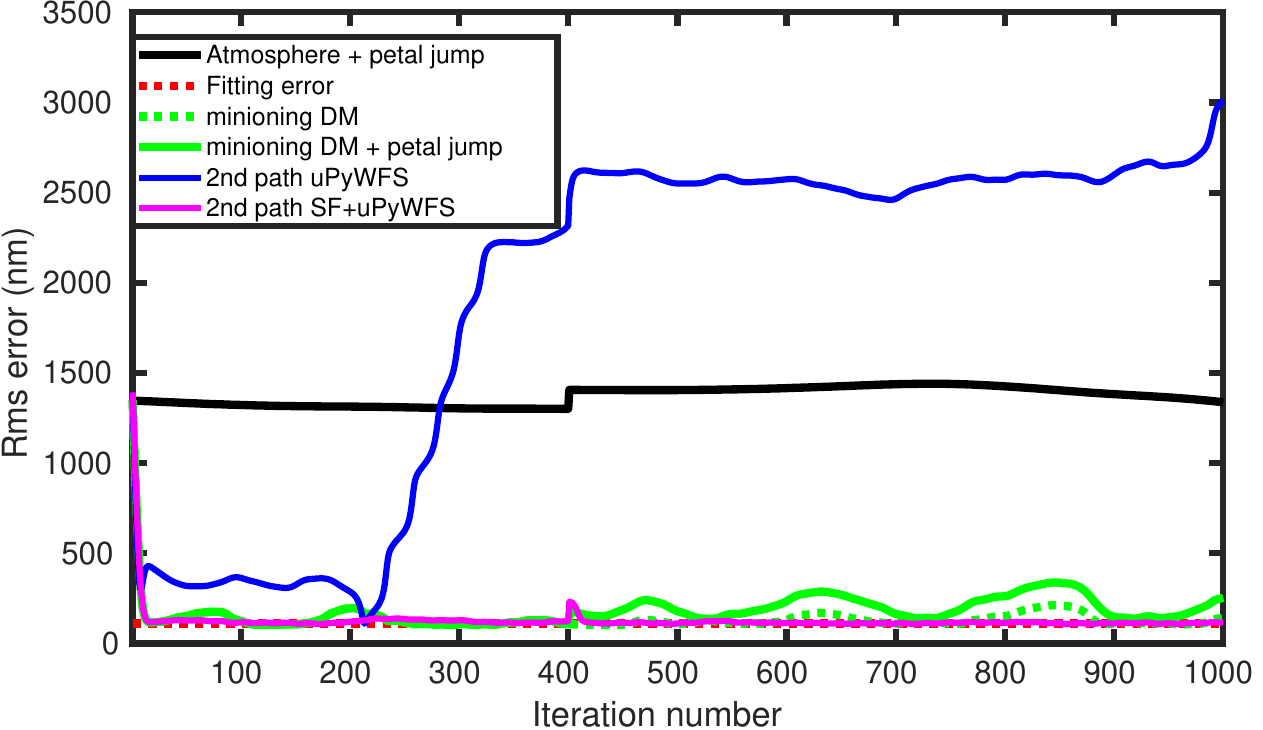}
    \caption{RMS error from atmosphere in black, first path AO residuals in green, first path assisted by uPyWFS-petalometer in blue and first path assisted by SF+uPyWFS-petalometer in magenta. A pure petal mode is added at frame 400 of 300nm PTV to simulate a petal mode exterior to the atmosphere. We see regular flares of the first path PyWFS while the SH+uPyWFS helps to keep the residual stable around zero petal during the whole sequence}
    \label{fig:2sensor_RMS}
\end{figure}

\begin{figure}
    \centering
    \includegraphics[width=\columnwidth]{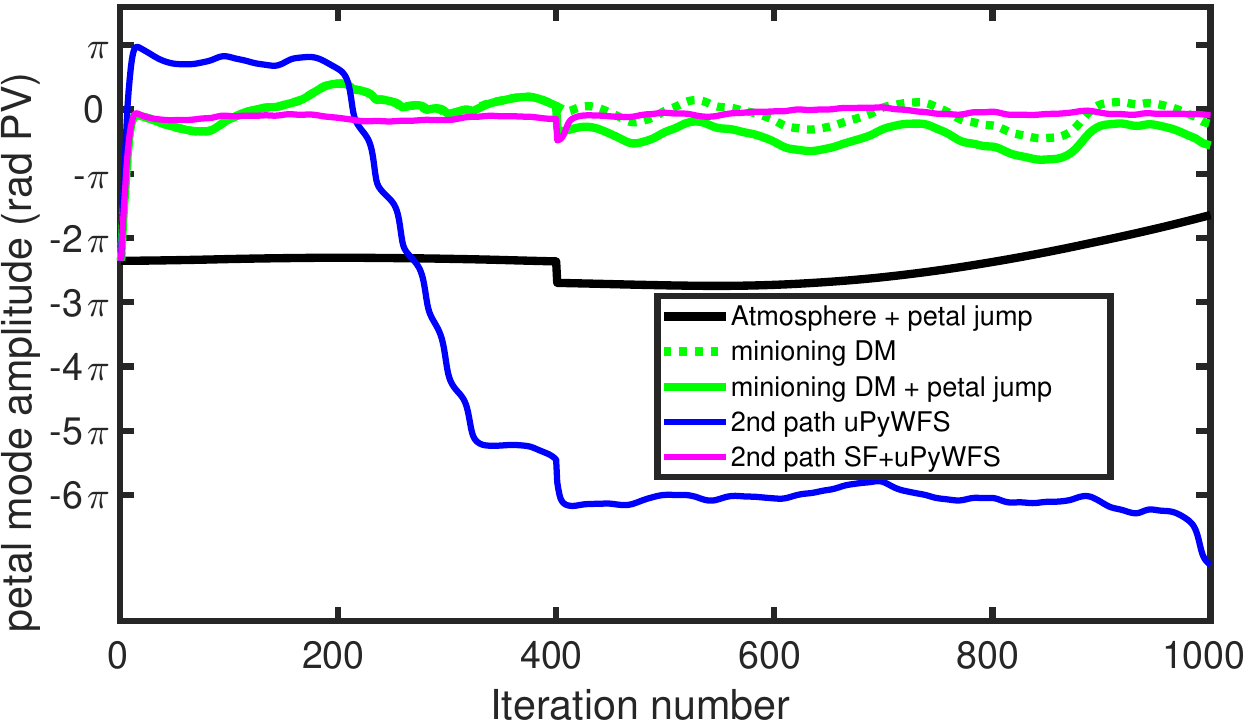}
    \caption{Residual phase of Fig. \ref{fig:2sensor_RMS} projected on petal mode. The first path creates petal mode residuals visible as oscillations between $-\pi$ and $+\pi$. uPyWFS keeps the residual petal closer to 0 petal, but with a jump of -6pi (around iteration 250) where it loses its locking of petal, while SF+uPyWFS stays locked around zero petal}
    \label{fig:2sensor_Petal}
\end{figure}

%\subsection{2 path system close loop}
Finally we simulate the full system of the $2$-path sensor to test the proposed concept of petalometer as presented in Fig.\ref{fig:2path_sensor}. 
We compare an uPyWFS and a SF+uPyWFS. The full system is described previously in Part $3$. The AO first path sensor (a modulated PyWFS) controls the DM (with simple Gaussian continuous influence functions, $20$x$20$ actuators), so it creates minioning type petal residuals. The petalometer commands a hypothetical DM with a pure petal mode as influence function. We subtract the atmospheric petal mode at frame $0$ from the atmospheric phase screen during the whole loop to start from a $0$ petal. Measuring the petal with a SF+uPyWFS is equivalent to using the pyramid in a full aperture gain mode (see \cite{plantetRevisitingComparisonShackHartmann2015}). With an optimized sensor, we can at least expect a flux distribution between AO-WFS and petal-WFS of 4000 / 6: number of modes used for AO, and number of petal modes. This would turn into a very small reduction (1/1000th in flux) in terms of system limit magnitude. On top of this, the characteristic times might be different between petal mode and atmosphere mode, depending on their origin. If the petal mode is slow enough (like LWE for instance), its measurement can be averaged with time on a few frames of AO loop, allowing to reduce furthermore the flux taken for the petalometer path.
 Since we are using monochromatic PyWFS the best possible result is to have petal mode locked at its initial value (zero) during the whole simulation. A bad petalometer would not be able to lock the petal mode at a stable value. 
Furthermore, we add (starting at frame $400$ = 0.4ms of simulation a $300$nm RMS static petal mode in the atmospheric phase. The aim is to test if our proposed strategy can measure and compensate for petal mode from other sources than the atmosphere. The parameters are the same as in Table \ref{table:AO_simconditions} and cover a $1s$ simulation. the results in RMS error and projected on petal mode are shown in Figs. \ref{fig:2sensor_RMS} and \ref{fig:2sensor_Petal} respectively

The result shows that our continuous influence functions allow the petal mode to stay around zero in the presence of petal flares. The residuals are mostly dominated by fitting error (dotted red line) with petal mode on top of it. But if an exterior petal mode is added during the simulation it is not measured and remains uncorrected as shown by the difference between the green and dotted green line (respectively with and without the $300$nm petal mode jump). The uPyWFS is not a good petalometer: due to the first path residuals it does not measure correctly the petal mode and jumps randomly of $2\pi$ petal mode value.
 We have shown that the SF+uPyWFS can measure correctly the petal mode surrounded by AO residuals, and use this measurement in a correction loop. Moreover, it is able to measure a sudden petal mode jump during the whole simulation. This $2$-path system allows the AO to stay as close to the fitting error as possible. 

\section{Conclusion}
When the 30m-Class telescopes will be completed and take their first scientific images, the petal mode will most certainly be an issue. Either for first light instrument where LWE is not controlled or for $2^{nd}$ generation instrument and in particular ExAO where the current petal is unacceptable.
 This paper proposed a new way to measure and control the petal mode in the loop using visible light. For this aim, we studied a $2$-path system with one sensor dedicated to atmospheric turbulence measurement and a second one dedicated to the measurement of the petal mode. 
 As a first proposed implementation we simulated a $2$ path system using a modulated PyWFS for the atmospheric control and an uPyWFS as the petalometer. We showed that the residuals of the first path still prevent the accurate reconstruction of the petal mode even by an uPyWFS. Another step is needed to reconstruct accurately the petal mode. By analysing the spatial structure of the residuals and the petal mode we showed that a focal plane spatial filter can improve significantly the petal mode reconstruction. After simulation and optimization of the spatial filter size the spatial filter seemed to greatly improve our reconstruction. With end-to-end simulation we confirmed its interest as a petalometer, precise enough to lock petal mode during the AO loop and capable of measuring unexpected petal flares.
The next steps are three fold. First the spatial filtering assisted reconstruction of petal mode should be tested on bench then on-sky using respectively the LOOPS bench at Laboratoire d'Astrophysique de Marseille and the PAPYRUS instrument at Observatoire de Haute Provence. 
Other potential petalometer solutions can be proposed and would surely benefit from the spatial filtering step as well (interferometric measurement with a adapted number of observables). Finally the solution must also be optimized to real system: ELT and GMT proposed SCAO systems to prepare for the second generation of instruments.

 \begin{acknowledgements} 
 This work benefited from support by the french government under the \emph{France 2030 investment plan, as part of the \emph{Initiative d'Excellence d'Aix-Marseille Université AMIDEX}, program number AMX-22-RE-AB-151,} support of the French National Research Agency (ANR) with WOLF (ANR-18-CE31-0018), APPLY (ANR-19-CE31-0011) and LabEx FOCUS (ANR-11-LABX-0013) the Programme Investissement Avenir F-CELT (ANR-21-ESRE-0008), the Action Spécifique Haute Résolution Angulaire (ASHRA) of CNRS/INSU co-funded by CNES, the ECOS-CONYCIT France-Chile cooperation (C20E02), the ORP H2020 Framework Programme of the European Commission’s (Grant number 101004719) and STIC AmSud (21-STIC-09),
\end{acknowledgements} 

\bibliography{aanda}

\begin{thebibliography}{22}
\providecommand{\natexlab}[1]{#1}
\providecommand{\url}[1]{\texttt{#1}}
\expandafter\ifx\csname urlstyle\endcsname\relax
  \providecommand{\doi}[1]{doi: #1}\else
  \providecommand{\doi}{doi: \begingroup \urlstyle{rm}\Url}\fi

\bibitem[{Bertrou-Cantou}(2021)]{bertrou-cantouValidationComposantsClefs2021}
A.~{Bertrou-Cantou}.
\newblock \emph{Validation Des Composants Clefs de l'optique Adaptative de
  Premi\`ere Lumi\`ere de l'instrument {{MICADO}} Pour l'{{ELT}}}.
\newblock These en pr\'eparation, Universit\'e de Paris (2019-....), 2021.

\bibitem[{Bertrou-Cantou} et~al.(2020){Bertrou-Cantou}, Gendron, Rousset,
  Ferreira, Sevin, Vidal, Cl{\'e}net, Buey, and
  Karkar]{bertrou-cantouPetalometryELTDealing2020}
A.~{Bertrou-Cantou}, E.~Gendron, G.~Rousset, F.~Ferreira, A.~Sevin, F.~Vidal,
  Y.~Cl{\'e}net, T.~Buey, and S.~Karkar.
\newblock Petalometry for the {{ELT}}: Dealing with the wavefront
  discontinuities induced by the telescope spider.
\newblock In \emph{Adaptive {{Optics Systems VII}}}, volume 11448, pages
  213--224. {SPIE}, December 2020.
\newblock \doi{10.1117/12.2562091}.

\bibitem[{Bertrou-Cantou} et~al.(2022){Bertrou-Cantou}, Gendron, Rousset, Deo,
  Ferreira, Sevin, and Vidal]{bertrou-cantouConfusionDifferentialPiston2022}
A.~{Bertrou-Cantou}, E.~Gendron, G.~Rousset, V.~Deo, F.~Ferreira, A.~Sevin, and
  F.~Vidal.
\newblock Confusion in differential piston measurement with the pyramid
  wavefront sensor.
\newblock \emph{Astronomy \& Astrophysics}, 658:\penalty0 A49, February 2022.
\newblock ISSN 0004-6361, 1432-0746.
\newblock \doi{10.1051/0004-6361/202141632}.

\bibitem[Bond et~al.(2022)Bond, Sauvage, Schwartz, Levraud, Chambouleyron,
  Correia, Fusco, and Neichel]{bondHARMONIELTWavefront2022}
C.~Bond, {\relax JF}.~Sauvage, N.~Schwartz, N.~Levraud, V.~Chambouleyron,
  C.~Correia, T.~Fusco, and B.~Neichel.
\newblock {{HARMONI}} at {{ELT}}: Wavefront control in {{SCAO}} mode.
\newblock In \emph{{{SPIE Astronomical Telescopes}} + {{Instrumentation}}}.
  {SPIE}, July 2022.
\newblock \doi{10.1117/12.2627713}.

\bibitem[Bonnefond et~al.(2016)Bonnefond, Tallon, Le~Louarn, and
  Madec]{bonnefondWavefrontReconstructionPupil2016}
S.~Bonnefond, M.~Tallon, M.~Le~Louarn, and {\relax PY}.~Madec.
\newblock Wavefront reconstruction with pupil fragmentation: Study of a simple
  case.
\newblock In E.~Marchetti, L.~Close, and {\relax JP}.~V{\'e}ran, editors,
  \emph{{{SPIE Astronomical Telescopes}} + {{Instrumentation}}}, page 990972,
  {Edinburgh, United Kingdom}, July 2016.
\newblock \doi{10.1117/12.2234034}.

\bibitem[Carlomagno et~al.(2020)Carlomagno, Delacroix, Absil, Cantalloube,
  Xivry, Pathak, Agocs, Bertram, Brandl, Burtscher, Doelman, Feldt, Glauser,
  Hippler, Kenworthy, Por, Snik, Stuik, and
  Boekel]{carlomagnoMETISHighcontrastImaging2020}
B.~Carlomagno, C.~Delacroix, O.~Absil, F.~Cantalloube, {\relax GO}.~Xivry,
  P.~Pathak, T.~Agocs, T.~Bertram, B.~Brandl, L~Burtscher, D.~Doelman,
  M.~Feldt, A.~Glauser, S.~Hippler, M.~Kenworthy, Emiel~H. Por, F.~Snik,
  R.~Stuik, and R.~Boekel.
\newblock {{METIS}} high-contrast imaging: Design and expected performance.
\newblock \emph{Journal of Astronomical Telescopes, Instruments, and Systems},
  6\penalty0 (3):\penalty0 035005, September 2020.
\newblock ISSN 2329-4124, 2329-4221.
\newblock \doi{10.1117/1.JATIS.6.3.035005}.

\bibitem[Cayrel(2012)]{cayrelEELTOptomechanicsOverview2012}
M.~Cayrel.
\newblock E-{{ELT}} optomechanics: Overview.
\newblock In L.~Stepp, R.~Gilmozzi, and H.~Hall, editors, \emph{{{SPIE
  Astronomical Telescopes}} + {{Instrumentation}}}, page 84441X, {Amsterdam,
  Netherlands}, September 2012.
\newblock \doi{10.1117/12.925175}.

\bibitem[Engler et~al.(2022)Engler, Louarn, V{\'e}rinaud, Weddell, and
  Clare]{englerFlipflopModulationMethod2022a}
B.~Engler, M.~Louarn, C.~V{\'e}rinaud, S.~Weddell, and R.~Clare.
\newblock Flip-flop modulation method used with a pyramid wavefront sensor to
  correct piston segmentation on {{ELTs}}.
\newblock \emph{Journal of Astronomical Telescopes, Instruments, and Systems},
  8\penalty0 (2):\penalty0 021502, March 2022.
\newblock ISSN 2329-4124, 2329-4221.
\newblock \doi{10.1117/1.JATIS.8.2.021502}.

\bibitem[Esposito et~al.(2003)Esposito, Pinna, Tozzi, Stefanini, and
  Devaney]{espositoCophasingSegmentedMirrors2003}
S.~Esposito, E.~Pinna, A.~Tozzi, P.~Stefanini, and N.~Devaney.
\newblock Cophasing of segmented mirrors using the pyramid sensor.
\newblock In \emph{Astronomical {{Adaptive Optics Systems}} and
  {{Applications}}}, volume 5169, pages 72--78. {SPIE}, December 2003.
\newblock \doi{10.1117/12.511507}.

\bibitem[Fauvarque(2017)]{fauvarqueOptimisationAnalyseursFront2017}
O.~Fauvarque.
\newblock \emph{Optimisation Des Analyseurs de Front d'onde \`a Filtrage
  Optique de {{Fourier}}}.
\newblock These de doctorat, Aix-Marseille, September 2017.

\bibitem[Fauvarque et~al.(2016)Fauvarque, Neichel, Fusco, Sauvage, and
  Girault]{fauvarqueGeneralFormalismFourierbased2016}
O.~Fauvarque, B.~Neichel, T.~Fusco, {\relax JF}.~Sauvage, and O~Girault.
\newblock General formalism for {{Fourier-based}} wave front sensing.
\newblock \emph{Optica}, 3\penalty0 (12):\penalty0 1440--1452, December 2016.
\newblock ISSN 2334-2536.
\newblock \doi{10.1364/OPTICA.3.001440}.

\bibitem[Haffert et~al.(2022)Haffert, Close, Hedglen, Males, Kautz, Bouchez,
  Demers, {Quiros-Pacheco}, Sitarski, Van~Gorkom, Long, Guyon, Schatz, Miller,
  Lumbres, Rodack, and Knight]{haffertHolographicDispersedFringe2022}
S.~Haffert, L.~Close, A.~Hedglen, J.~Males, M.~Kautz, A.~Bouchez, R.~Demers,
  F.~{Quiros-Pacheco}, B.~Sitarski, K.~Van~Gorkom, J.~Long, O.~Guyon,
  L.~Schatz, K.~Miller, J.~Lumbres, A.~Rodack, and J.~Knight.
\newblock The {{Holographic Dispersed Fringe Sensors}} ({{HDFS}}): Phasing the
  {{Giant Magellan Telescope}}, June 2022.

\bibitem[Hippler et~al.(2019)Hippler, Feldt, Bertram, Brandner, Cantalloube,
  Carlomagno, Absil, Obereder, Shatokhina, and
  Stuik]{hipplerSingleConjugateAdaptive2019}
S.~Hippler, M.~Feldt, T.~Bertram, W.~Brandner, F.~Cantalloube, B.~Carlomagno,
  O.~Absil, A.~Obereder, J.~Shatokhina, and R.~Stuik.
\newblock Single conjugate adaptive optics for the {{ELT}} instrument
  {{METIS}}.
\newblock \emph{Experimental Astronomy}, April 2019.
\newblock \doi{10.1007/s10686-018-9609-y}.

\bibitem[Leboulleux et~al.(2022)Leboulleux, Carlotti, N'Diaye, Cantalloube,
  Milli, {Bertrou-Cantou}, Mouillet, Pourr{\'e}, and
  V{\'e}rinaud]{leboulleuxRedundantApodizedPupils2022}
L~Leboulleux, A.~Carlotti, M.~N'Diaye, F.~Cantalloube, J.~Milli,
  A.~{Bertrou-Cantou}, D.~Mouillet, N.~Pourr{\'e}, and C.~V{\'e}rinaud.
\newblock Redundant {{Apodized Pupils}} ({{RAP}}) for high-contrast imagers
  robust to segmentation-due aberrations and island effects.
\newblock In \emph{Advances in {{Optical}} and {{Mechanical Technologies}} for
  {{Telescopes}} and {{Instrumentation V}}}, page~63, August 2022.
\newblock \doi{10.1117/12.2629276}.

\bibitem[Martins et~al.(2022)Martins, Holzl{\"o}hner, V{\'e}rinaud, and
  Kleinclaus]{martinsTransientWavefrontError2022}
D.~Martins, R.~Holzl{\"o}hner, C.~V{\'e}rinaud, and C.~Kleinclaus.
\newblock Transient wavefront error from cooled air downwind of telescope
  spiders.
\newblock In G.~Angeli and P.~Dierickx, editors, \emph{Modeling, {{Systems
  Engineering}}, and {{Project Management}} for {{Astronomy X}}}, page~70,
  {Montr\'eal, Canada}, August 2022. {SPIE}.
\newblock ISBN 978-1-5106-5355-9 978-1-5106-5356-6.
\newblock \doi{10.1117/12.2627683}.

\bibitem[Plantet et~al.(2015)Plantet, Meimon, Conan, and
  Fusco]{plantetRevisitingComparisonShackHartmann2015}
C.~Plantet, S.~Meimon, J.-M. Conan, and T.~Fusco.
\newblock Revisiting the comparison between the {{Shack-Hartmann}} and the
  pyramid wavefront sensors via the {{Fisher}} information matrix.
\newblock \emph{Optics Express}, 23\penalty0 (22):\penalty0 28619--28633,
  November 2015.
\newblock ISSN 1094-4087.
\newblock \doi{10.1364/OE.23.028619}.

\bibitem[Ragazzoni(1996)]{ragazzoniPupilPlaneWavefront1996a}
R.~Ragazzoni.
\newblock Pupil plane wavefront sensing with an oscillating prism.
\newblock \emph{Journal of Modern Optics}, 43\penalty0 (2):\penalty0 289--293,
  February 1996.
\newblock ISSN 0950-0340.
\newblock \doi{10.1080/09500349608232742}.

\bibitem[Sauvage et~al.(2015)Sauvage, Fusco, Guesalaga, Wizinowitch, O'Neal,
  N'Diaye, Vigan, Grard, Lesur, Mouillet, Beuzit, Kasper, Le~Louarn, Milli,
  Dohlen, Neichel, Bourget, Heigenauer, and Mawet]{sauvageLowWindEffect2015}
{\relax JF}.~Sauvage, T.~Fusco, A.~Guesalaga, P.~Wizinowitch, J.~O'Neal,
  M.~N'Diaye, A.~Vigan, J.~Grard, G.~Lesur, D.~Mouillet, {\relax JL}.~Beuzit,
  M.~Kasper, M.~Le~Louarn, J.~Milli, K.~Dohlen, B.~Neichel, P.~Bourget,
  P.~Heigenauer, and D.~Mawet.
\newblock Low {{Wind Effect}}, the main limitation of the {{SPHERE}}
  instrument.
\newblock 2015.
\newblock \doi{10.20353/K3T4CP1131541}.

\bibitem[Schwartz et~al.(2017)Schwartz, Sauvage, Correia, Petit,
  {Quiros-Pacheco}, Fusco, Dohlen, El~Hadi, Thatte, Clarke, Paufique, and
  Vernet]{schwartzSensingControlSegmented2017}
N.~Schwartz, {\relax JF}.~Sauvage, C.~Correia, C.~Petit, F.~{Quiros-Pacheco},
  T.~Fusco, K.~Dohlen, K.~El~Hadi, N.~Thatte, F.~Clarke, J.~Paufique, and
  J.~Vernet.
\newblock Sensing and control of segmented mirrors with a pyramid wavefront
  sensor in the presence of spiders.
\newblock In \emph{Proceedings of the {{Adaptive Optics}} for {{Extremely Large
  Telescopes}} 5}. {Instituto de Astrof\'isica de Canarias (IAC)}, 2017.
\newblock \doi{10.26698/AO4ELT5.0015}.

\bibitem[Schwartz et~al.(2018)Schwartz, Sauvage, Correia, Neichel, Fusco,
  {Quiros-Pacheco}, Dohlen, El~Hadi, Agapito, Thatte, and
  Clarke]{schwartzAnalysisMitigationPupil2018}
N.~Schwartz, {\relax JF}.~Sauvage, C.~Correia, B.~Neichel, T.~Fusco,
  F.~{Quiros-Pacheco}, K.~Dohlen, K.~El~Hadi, G.~Agapito, N.~Thatte, and
  F.~Clarke.
\newblock Analysis and mitigation of pupil discontinuities on adaptive optics
  performance.
\newblock In D.~Schmidt, L.~Schreiber, and L.~Close, editors, \emph{Adaptive
  {{Optics Systems VI}}}, page~75, {Austin, United States}, July 2018. {SPIE}.
\newblock ISBN 978-1-5106-1959-3 978-1-5106-1960-9.
\newblock \doi{10.1117/12.2313129}.

\bibitem[Usuda et~al.(2014)Usuda, Ezaki, Kawaguchi, Nagae, Kato, Takaki,
  Hirano, Hattori, Tabata, Horiuchi, Saruta, Sofuku, Itoh, Oshima, Takanezawa,
  Endo, Inatani, Iye, Sadjadpour, Sirota, Roberts, and
  Stepp]{usudaPreliminaryDesignStudy2014}
T.~Usuda, Y.~Ezaki, N.~Kawaguchi, K.~Nagae, A.~Kato, J.~Takaki, M.~Hirano,
  T.~Hattori, M.~Tabata, Y.~Horiuchi, Y.~Saruta, S.~Sofuku, N.~Itoh, T.~Oshima,
  T.~Takanezawa, M.~Endo, J.~Inatani, M.~Iye, A.~Sadjadpour, M.~Sirota,
  S.~Roberts, and L.~Stepp.
\newblock Preliminary design study of the {{TMT Telescope}} structure system:
  Overview.
\newblock In L.~Stepp, R.~Gilmozzi, and H.~Hall, editors, \emph{{{SPIE
  Astronomical Telescopes}} + {{Instrumentation}}}, page 91452F, {Montr\'eal,
  Quebec, Canada}, July 2014.
\newblock \doi{10.1117/12.2055767}.

\bibitem[V{\'e}rinaud(2004)]{verinaudNatureMeasurementsProvided2004a}
C.~V{\'e}rinaud.
\newblock On the nature of the measurements provided by a pyramid wave-front
  sensor.
\newblock \emph{Optics Communications}, 233\penalty0 (1):\penalty0 27--38,
  March 2004.
\newblock ISSN 0030-4018.
\newblock \doi{10.1016/j.optcom.2004.01.038}.

\end{thebibliography}
\bibliographystyle{plainnat}

\end{document}